\documentstyle[pra,aps,epsf,amssymb]{revtex}

\newcommand{\bra}[1]{\mathop{\langle\left.{#1}\right|}\nolimits}
\newcommand{\ket}[1]{\mathop{\left|{#1}\right.\rangle}\nolimits}
\newcommand{\avr}[1]{\mathop{\langle{#1}\rangle}\nolimits}

\begin{document}
\draft


\title{Coherent and compatible information: a basis to information
analysis\\ of quantum systems}
\author{B.\ A.\ Grishanin\thanks{grishan@comsim1.phys.msu.su} and
V.\ N.\ Zadkov\thanks{zadkov@comsim1.phys.msu.su}}
\address{International Laser Center and Department of Physics\\
M.\ V.\ Lomonosov Moscow State University, 119899 Moscow, Russia}
\date{August 8, 2001}
\maketitle

\begin{abstract}
Relevance of key quantum information measures for analysis of quantum systems
is discussed. It is argued that possible ways of measuring quantum information
are based on compatibility/incompatibility of the quantum states of a quantum
system, resulting in the coherent information and introduced here the
compatible information measures, respectively. A sketch of an information
optimization of a quantum experimental setup is proposed.
\end{abstract}
\pacs{PACS numbers: 03.65.Bz, 03.65.-w, 89.70.+c}

\section{Introduction}
\label{section:intro}

The field of quantum information was born at the same time the basic laws of
quantum physics had been established and since that time it plays an important
role in physics. One could even say that quantum information theory was
established prior the classical Shannon information theory. In favour of this,
Bloch interpretation of the wave function or information meaning of the quantum
collapse postulate could be mentioned \cite{sudbery}. Moreover, any quantum
effect, i.e., essentially microscopic process of atom's spontaneous emission or
macroscopic superconductivity transition, is associated with the corresponding
process of quantum information transmission. Although importance of the quantum
information concept was recognized long ago, not much attention has been paid
to its practical importance until now, when modern experiments in quantum
optics provide detailed control over quantum states of quantum systems. This
allow us not only to think about quantum information as of an abstract concept,
but apply it to real quantum systems and real experiments.

Sometimes it is expostulated that in physics one should necessarily deal with
physical values, and if dealing only with physical states it is not physics but
mathematics. Yet it is not true---whenever the states are specified as the
states of a physical model, they provide physically meaning information. As an
example, let us discuss an operator $\hat A$ in Hilbert space $H$ as a
representation of a physical variable. Then, writing $\hat A$ as a spectral
decomposition $\hat A= \sum\lambda_n\ket n\bra n$ we represent it with two
types of mathematical objects: $\lambda_n$, the possible physical values, and
$\ket n$, the corresponding physical states. The latter contain the most
general type of physical information on physical events regardless of the
values $\lambda_n$.

The most general concept of classical information is the information theory
introduced by Shannon\cite{shannon49,gallagher68}. This very elegant theory is
based on the specific property of classical ensembles, which follows from the
basic principles of quantum physics. This property is the
\emph{reproducibility} of classical events: statistically there is no
difference either you have at input and output physically the same system or
its informationally equivalent copies. The latter case is impossible in quantum
world, which gives a rise to a discussion whether the Shannon approach can be
applicable to the quantum systems or not\cite{brukner1,Hall2,brukner3}. As we
will show, the traditional Shannon entropy and information measures can be
successfully used for analysis of quantum systems, if correctly applied with
clear understanding of the basic differences between the classical and quantum
states ensembles.

\begin{figure}
\begin{center}
\epsfxsize=0.25\textwidth\epsfclipon\leavevmode\epsffile{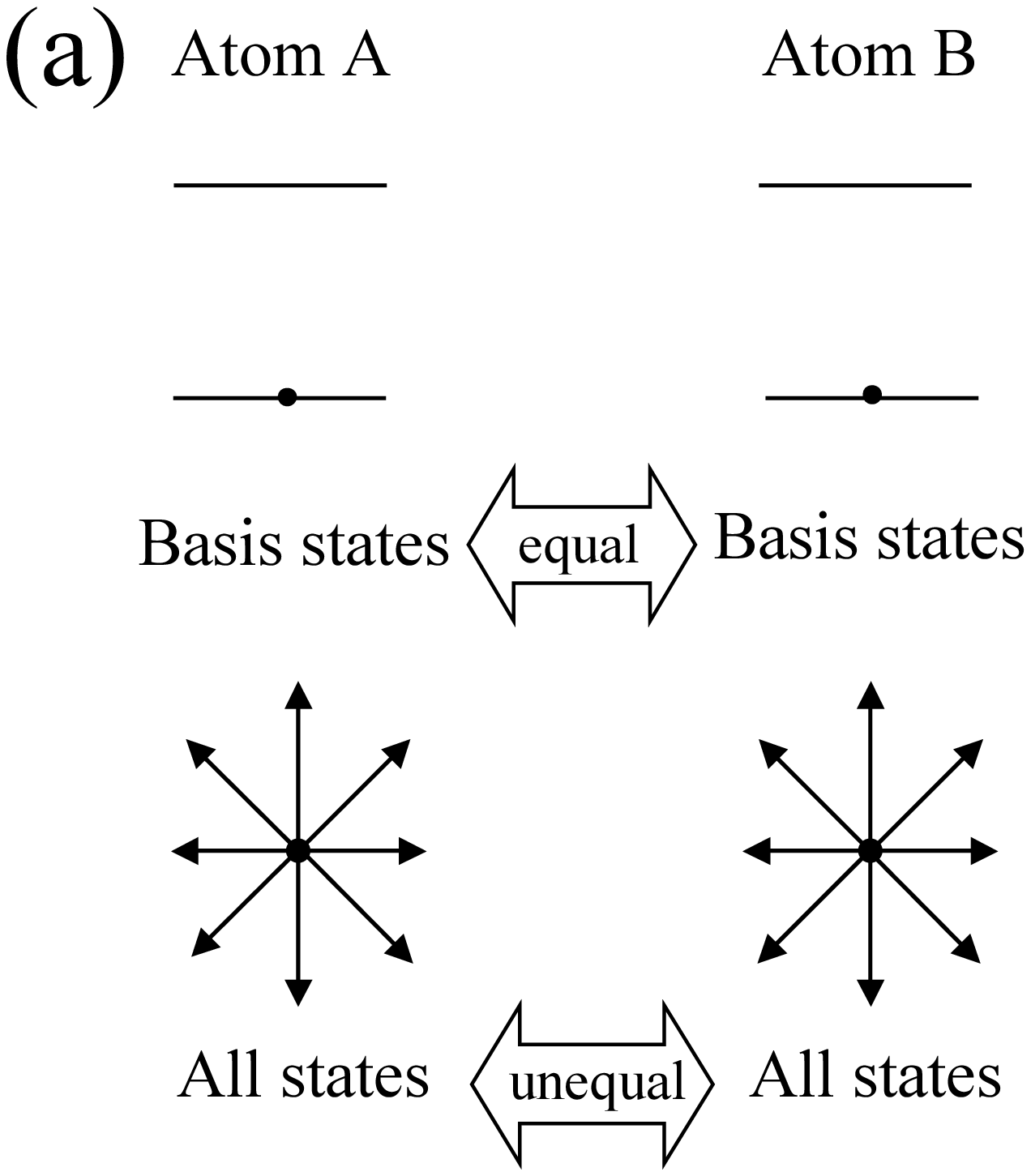}\hspace{1.5cm}
\epsfxsize=0.25\textwidth\epsfclipon\leavevmode\epsffile{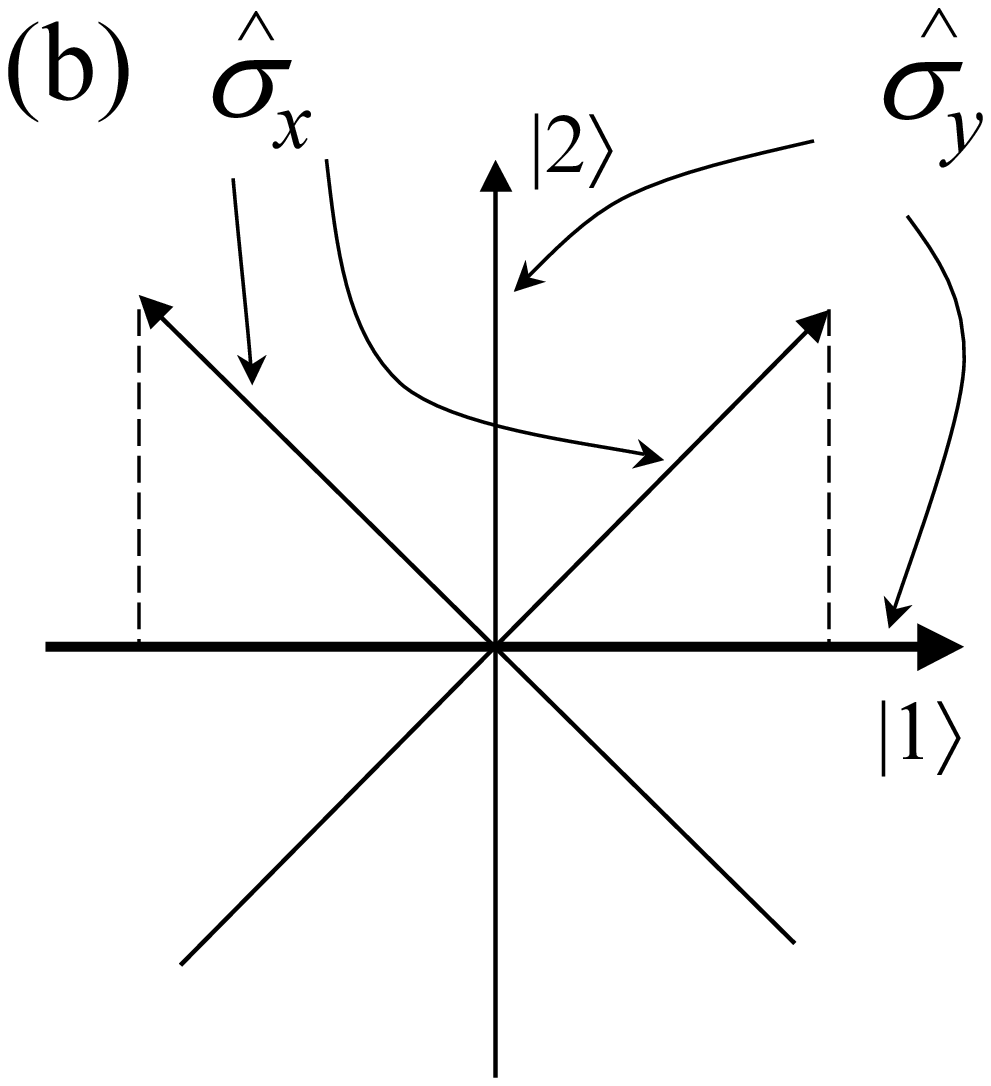}
\end{center}
\caption{a) Equivalence of compatible basis-states
ensembles and inequivalence of incompatible all-states ensembles of two
two-level atoms. b) Vacuum fluctuations as a result of incompatibility: eigen
states of $\hat\sigma_x$ have equal non-zero probabilities $p_\pm=1/2$ at the
eigen atomic state $\ket1$, thus providing nonzero fluctuations.}\label{fig1}
\end{figure}

Let us discuss, for example, two atoms in the same state (Fig.\ \ref{fig1}a).
Term the ``same" needs to be refined for the case of quantum systems, by
contrast with its classical meaning. In the classical case, we take into
account only two basis states of each atom. Then, we are free to suppose that
either these basic states correspond to two different atoms or to one and the
same atom. Important is that there is only one non-zero probability state in a
combined system of two atoms---if a state of one of the two considered atoms is
given, another atom has a non-zero probability state. In quantum case, two
atoms have additional states with non-zero probability due to the internal
quantum uncertainty (Fig.\ \ref{fig1}b). It is well known that this uncertainty
results for a harmonic oscillator in vacuum fluctuation energy $\hbar\omega/2$.
In our case of two-level atoms it takes the form of the non-zero values
$\hat\sigma_x^2=\hat\sigma_y^2= \hat I$, where Pauli matrices $\hat
\sigma_{x,y}$ are treated as cosine and sine amplitudes of the atomic
oscillator. The corresponding fluctuations are different for these two atoms,
notwithstanding the latter are in the ``same" state, which belongs to different
atoms possessing individual internal incompatible ensembles of quantum states.
Indeed, the average squared differences $\bigl(\hat\sigma_x^A-\hat\sigma_x^B
\bigr)^2$, $\bigl(\hat\sigma_y^A -\hat\sigma_y^B\bigr)^2$ are both different
from zero due to the non-commutativity of their operators with the population
operators $\hat\sigma_z$, the latter yield certainly zero difference
$\hat\sigma_z^A-\hat\sigma_z^B$.

What we can learn from the considered above example is that when ensembles of
quantum system states are incompatible, i.e., non-orthogonal states of each
atom (eigen states of the corresponding non-commuting operators) are involved,
the states of two different atoms are always different with respect to all
their ever coexisting internal quantum states allowed by the quantum
uncertainty.

This statement can be expressed in a quantitative form as strict positivity of
the average operator of the squared difference between the ortho-projectors
onto the  corresponding wave functions of the two atoms:
\begin{displaymath}
\hat\varepsilon=\int
\Bigl(\ket\alpha\bra\alpha\otimes\hat I_B-\hat I_A\otimes\ket\alpha
\bra\alpha\Bigr)^2\frac{dV_\alpha}{D} =\ket{\ket0}\bra{\bra0}+\frac{1}{3}
\sum\limits_{k=1}^3\ket{\ket{k}}\bra{\bra{k}}\geqq\frac{1}{3},
\end{displaymath}

\noindent where integration is made over the Bloch sphere of the states
$\ket\alpha$ with the volume differential $dV_\alpha=\sin \vartheta\,d\vartheta
d\varphi/(2\pi)$ and the total volume $V_\alpha{=}D{=}2$. This bipartite
operator has two eigen subspaces composed of a singlet and triplet Bell states
$\ket{\ket{k}}$, corresponding to the eigen squared difference values
$\varepsilon_k=1,1/3$, the singlet one being three times bigger.

At this point, one can conclude that the key difference between classical and
quantum information lies in \emph{compatibility} or \emph{incompatibility} of
the states associated with the information of interest. The one-time states of
different systems are always compatible. Therefore, they cannot copy one
another if states of each system include internally incompatible states.
Conversely, two-time states of the deterministically transformed system are
always incompatible. Two-time states of different systems can be either
compatible or not.

In this paper, we will classify quantum information in connection with the
compatibility property described above. In this vein, we can distinguish
four main types of information listed below:
\begin{itemize}
\item \emph{Classical information}---all the states are compatible and in
original form of information theory quantum systems are not
discussed\cite{shannon49,gallagher68}. Note that classical information can be
well transmitted through the quantum channels and also can be of interest in
Quantum Physics.

\item \emph{Semiclassical information}---all the input information is
given by classical states $\lambda$ and the output states include internal
incompatibility in the form of all states of a Hilbert space $H$, which are
automatically compatible with the input states. The quantum channel is
generally described via a classical parameter dependent on the ensemble of
mixed states $\hat\rho_\lambda$\cite{holevo73,hall97}.

\item \emph{Coherent information}---both input and output are spaces
composed of internally incompatible states, plus these spaces are also
incompatible and connected via a channel superoperator $\cal N$ transforming
the input density matrix into the output one: $\hat\rho_B={\cal
N}\hat\rho_A$\cite{schumacher96,lloyd97}. It is a ``flow" of quantum
incompatibility from one system to another.

\item \emph{Compatible information}---both input and output are composed of
internally incompatible states, which are mutually compatible.
\end{itemize}

While three first types of information where thoroughly discussed in the
literature\cite{gallagher68,holevo73,schumacher96}, including the recently
introduced coherent information measure, the compatible information is
introduced here for the first time. This new type of quantum information is
defined for a compound bipartite quantum system with the compatible input and
output, which include internal quantum incompatibility.

In our view, the coherent and compatible information exhaust all possible
qualitatively different types of information in quantum channels. Presented in
the paper feasibility analysis of using these two measures of information for
information analysis of real experimental schemes shows that only compatible
information turns to be suitable for information effectiveness analysis of an
experimental scheme (in the following we will simply call an experimental
scheme an ``experimental setup").

\section{Coherent Information}

\subsection{Physical meaning of coherent information}
\label{section:meaning}

The coherent information quantitatively represents an amount of incompatible
information, which is transferred from one space to another. A case of one and
the same space can be considered, as well. A trivial case of the coherent
information exchange is a dynamic evolution represented with the unitary time
evolution operator $U$, $\hat\rho_B=U\hat\rho_AU^{-1}$. Then, all pure states
$\psi$ allowed by the initial density matrix $\hat\rho_A$ are transformed with
no distortion, and the transmitted coherent information coincides with its
initial amount. The latter is measured, by definition, with the von Neumann
entropy, which reads
\begin{equation}\label{Ic0}
I_c=S[\hat\rho_B]=S[\hat\rho_A]=-{\rm Tr}\,\hat\rho_A\log\hat\rho_A.
\end{equation}

\noindent This definition yet demands additional justifying in terms revealing
an operational meaning of the density matrix, which is given in a
self-consistent quantum theory as a result of averaging of a pure state in a
compound system over the auxiliary variables. Then, Eq.~(\ref{Ic0}) describes
an entanglement of the input system $A$ with a \emph{reference} system $R$,
which corresponds to a proper pure state $\Psi_{AR}$, ${\rm Tr}_R
\ket{\Psi_{AR}}\bra{\Psi_{AR}}=\hat\rho_A$ of a combined $A{+}R$ system. Thus
quantitative measuring of the coherent information is done in terms of the
mutually compatible states of two different systems, $A$ and $R$, while
information transfers from input $A$ to the output $B$, which differs from $A$
here only with a unitary transformation.

\begin{figure}
\begin{center}
\epsfxsize=0.3\textwidth\epsfclipon\leavevmode\epsffile{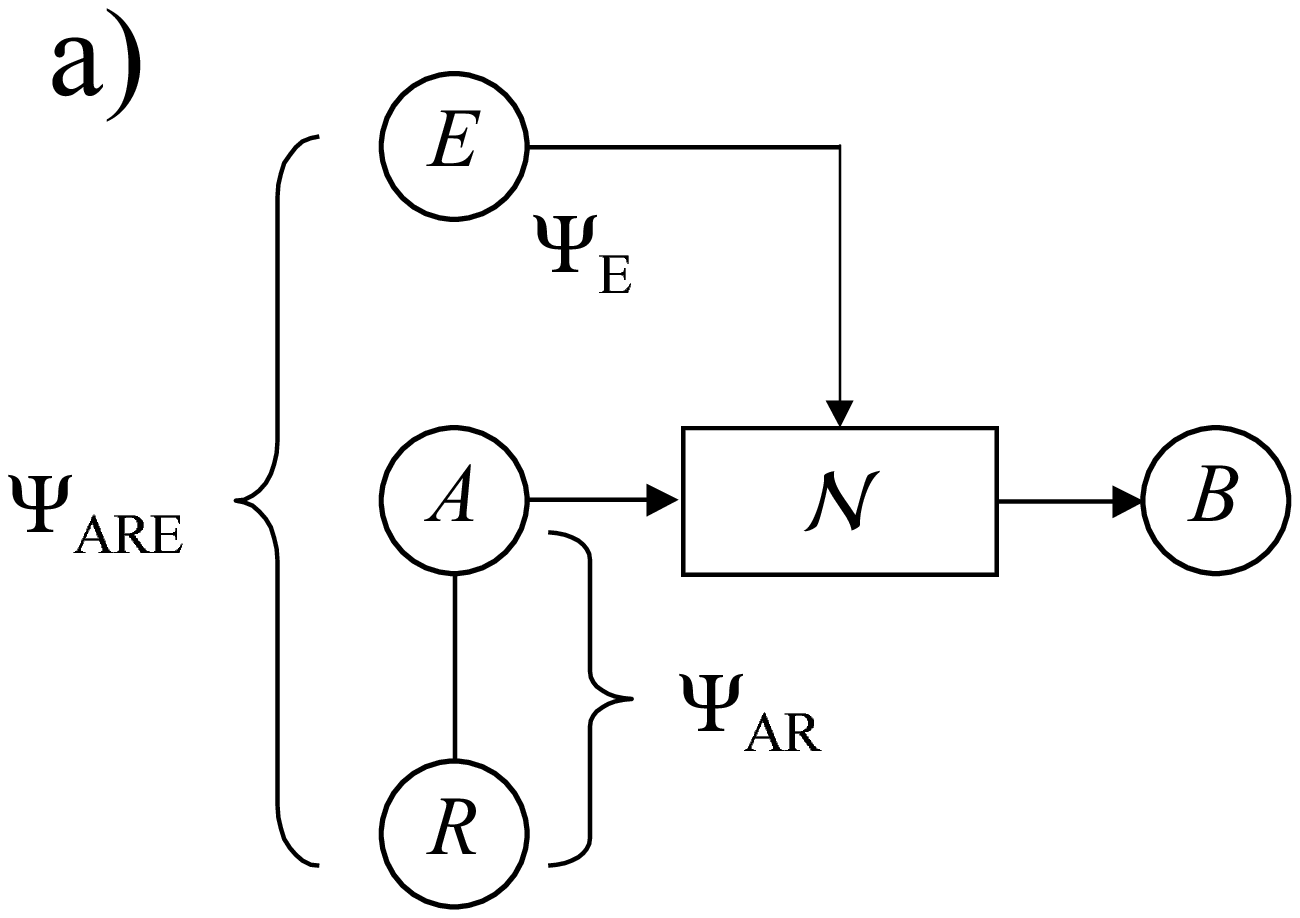}\hspace*{1cm}
\epsfxsize=0.4\textwidth\epsfclipon\leavevmode\epsffile{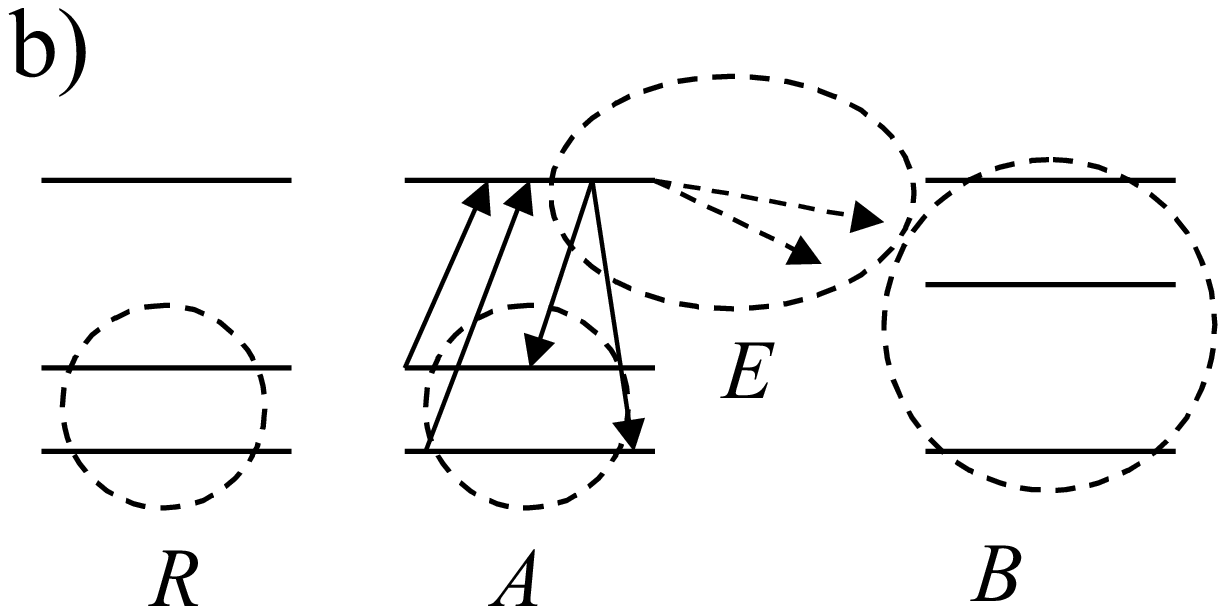}
\end{center}
\caption{a) The most general scheme of quantum information system,
composed of input $A$, reference system $R$, channel $\cal N$ with
noisy environment $E$, and output $B$. b) An example of physical
implementation of a quantum information system: an input $A$ and a
reference system $R$ are the ground states of two entangled atomic
$\Lambda$-systems, information channel $\cal N$ is provided with the
laser excitation of an input system to the radiative upper level, the
two photon field states corresponding to the emitted photons together
with the vacuum state provide an output $B$, and all other field states
together with the excited atomic state form the environment
$E$.}\label{fig2}
\end{figure}

To complete the general structure of the information system, an information
channel $\cal N$ with the attached noisy {\em environment} $E$ should be
added (Fig.\ \ref{fig2}a)\cite{schumacher98}.

The definition of the coherent information for a general type of channel reads
as\cite{schumacher98}
\begin{equation}\label{Ic}
I_c=S[\hat\rho_B]-S\bigl[({\cal N}\otimes{\cal I})\ket{\Psi_{AR}}
\bra{\Psi_{AR}}\bigr],
\end{equation}

\noindent where $\cal I$ is the identical superoperator applied to the
variables of the reference system. The second term is the \emph{entropy
exchange}, which is non-zero due to the exchange between the subsystems $A{+}R$
and $E$, which is when ${\cal N}\neq{\cal I}$. Channel superoperator $\cal N$
transforms the states of input $A$ according to the equation
\begin{equation}\label{N}
\hat\rho_B={\cal N}\hat\rho_A={\rm Tr}_R({\cal N}\otimes{\cal I})
\ket{\Psi_{AR}}\bra{\Psi_{AR}}
\end{equation}

\noindent to the states of output $B$, which is again compatible with the
reference system $R$ because of no entanglement between them at this
transformation. A physical meaning of Eq.~(\ref{Ic}) is switched then from an
incompatibility flow to a specific measure for a {\em preserved} entanglement
between the \emph{compatible} systems $R$ and $B$, which is left after
transmission through the channel. In a general case, output $B$ may be
physically different from $A$ and even represented with a Hilbert space of
different structure, $H_B\neq H_A$ \cite{PRA00,lasphys00}, as shown for a
specific example of a physical information system in Fig.\ \ref{fig2}b.

Now we will try to answer a question how the coherent information measure can
be used in physics? Quantum theory is usually applied to the calculation of
some average values $\avr{\hat A}=\sum\lambda_n\avr{\ket{n}\!\bra{n}}$, where
$\lambda_n$ and $\ket{n}$ denote the eigen values and eigen vectors of an
operator $\hat A$. This expansion represents averaging of physical variables in
terms of probabilities $P_n^A=\avr{\ket{n}\!\bra{n}}$ of quantum states
$\ket{n}$. As far as there is an innumerable set of all possible variables and
it is much richer than the set of all quantum states, description of the
correspondences between the physical states, apart of physical values, provides
a more general information on the physical correspondences the most economical
way. Laws for coherent information exchange follow the most basic laws of
quantum physics, as they show the most general features of interaction between
two systems of interest chosen as input and output and connected with a
one-to-one transformation of the input states. In fact, the dependencies of the
coherent information on the system parameters are even more basic than those of
specific physical values.

Let us consider, for example, a Dicke problem for which an information exchange
shows the same oscillation type of dynamics as the energy exchange between the
two atoms, assisted with the radiation damping \cite{PRA00}. This oscillatory
evolution is characteristic not only for the energy, but also for many other
variables. Therefore, there is a point in considering evolution of the coherent
information instead of working with many other variables. One should also keep
in mind the physical meaning of the coherent information as a preserved
entanglement. The latter, in its turn, is a characteristic of an internal
incompatibility exchange between the mutually compatible sets of states for the
reference and input systems, $H_R$ and $H_A$. Among other types of quantum
information the coherent information distinguishes between two types of
information, corresponding to the exchange via classical information and
quantum entanglement. The coherent information is nonzero only for the latter
case. Thus, it is adequate to discuss how well the given information
transmission channel preserves the capability of using the output as an
equivalent of the input to realize a task, when quantum properties of a signal
are essential. This problem received much attention in the literature (see
Ref.\ \onlinecite{QC} and references therein).

One can also be interested in applying the coherent information concept to an
analysis of a specific model of a quantum channel. One of the examples is
discussed in Sec.~\ref{section:rate}.

\subsection{One-time coherent information}
\label{section:one-time}

A first step towards information characterization of a two-side quantum
channel could be undertaken by formal quantum generalization of the
classical Shannon mutual information $I=S_A+S_B-S_{AB}$:
\begin{equation}\label{Istr}
I=S[\hat\rho_A]+ S[\hat\rho_B]- S[\hat \rho_{AB}],
\end{equation}

\noindent which is valid if the joint density matrix $\hat\rho_{AB}$ is given
and treated as a strict analogue of classical joint probability distribution
$P_{AB}$ \cite{rls65}. Evidently, to apply this formula to quantum systems we
should suppose that $A$ and $B$ states are mutually compatible, which is valid
for the one-time states of the corresponding physical systems, unless they
belong to the same system, both as input and output. Note at this point that
physical meaning of $I$ still remains unclear\cite{lindblad91,holevo98}. It
could be clarified by taking into account striking difference between the
classical and quantum information channels. Generally, as it follows from
Eq.~(\ref{N}), quantum input and output are incompatible, being taken for a
single system at two time instants. Thus, $A$ and $B$ cannot be treated as
input and output, and their further specification must be made for the quantum
case.

\begin{figure}[ht]
\begin{center}
\epsfxsize=0.35\textwidth\epsfclipon\leavevmode\epsffile{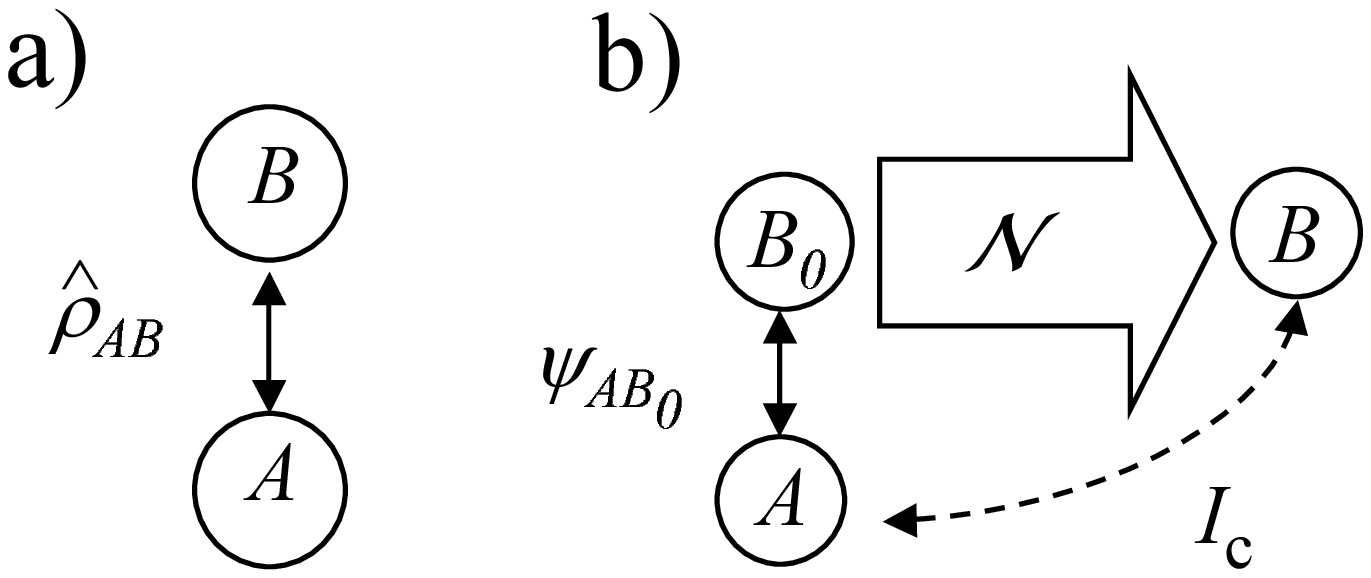}
\end{center}
\caption{Reconstruction of the quantum information system corresponding
to the given joint density matrix $\hat\rho_{AB}$: mathematical
description of a channel providing one-time coherent information (a)
and correspondence with the Schumacher's treatment (b).}\label{fig3}
\end{figure}

Let us then specify $A$ as the reference system and $B$ as the output for a given
joint density matrix $\hat\rho_{AB}$ as it is shown in Fig.\ \ref{fig3}. The
input $B_0$ and the channel $\cal N$ are not introduced explicitly but through
their action, resulting in the given density matrix $\hat\rho_{AB}$.

\noindent The pure state $\Psi_{AB_0}$ of the input--reference system and the
channel superoperator $\cal N$ should obey the equation
\begin{equation}\label{rhoAB}
\hat\rho_{AB}=\bigl({\cal I}{\otimes}{\cal
N}\bigr)\ket{\Psi_{AB_0}}\bra{\Psi_{AB_0}}.
\end{equation}

\noindent This automatically provides the coincidence of the partial density
matrix of the reference state
\begin{displaymath}\hat\rho_A={\rm Tr}_{B_0}\ket{\Psi_{AB_0}}
\bra{\Psi_{AB_0}}
\end{displaymath}

\noindent with the partial density matrix $\hat\rho_A={\rm Tr}_B\hat\rho_{AB}$
calculated by averaging of the given $A{+}B$ state, as far as trace over $B_0$
of Eq.~(\ref{rhoAB}) is invariant on $\cal N$.

Then, the corresponding {\em one-time} coherent information can be defined as
\begin{equation}\label{I1t}
I_c=S[\hat\rho_B]-S[\hat\rho_{AB}],
\end{equation}

\noindent which by contrast with the quantity (\ref{Istr}) lacks the term
$S[\hat\rho_A]$. Term ``one-time" here may not have in general case a strict
meaning, because any two compatible quantum systems $A$ and $B$, even related
to different time instants, can be treated as related to the one time instant
after the corresponding transformation of states.

Additional property of one-time coherent information is that definition
(\ref{I1t}) lacks symmetry by contrast with (\ref{Istr}). Moreover, the
coherent information can be negative. The latter is evident for the density
matrices $\hat\rho_{AB}$ corresponding to the purely classical information
exchange via orthogonal bases, $\hat\rho_{AB}= \sum P_{ij}\ket{i}\ket{j}
\bra{j}\bra{i}$. Then, the entropies reduce to the classical entropies
$S[\hat\rho_{AB}]{=}S_{AB}{=}-\sum P_{ij}\log P_{ij}$, $S[\hat\rho_B]{=}S_B{=}
-\sum P_j\log P_j$ and $S_{AB}{>}S_B$. Negative value of the coherent
information means that the entropy exchange prevails information transmission,
so it is reasonable to set $I_c=0$ in this case.

\subsection{Coherent information exchange rate in the $\Lambda$-system}
\label{section:rate}

Information system presented in Fig.\ \ref{fig2}b plays a special role in new
applications based on nonclassical properties of quantum information, e.g.,
quantum cryptography and quantum computations. Key elements in such
applications are atomic $\Lambda$-systems, which thought to be promising
elements (qubits) to store quantum information and are convenient to manipulate
with the help of laser radiation\cite{QC,UFN01}. For our system
(Fig.\ \ref{fig2}b) treating second $\Lambda$-system as a reference system has a
reasonable justification, as the entanglement of two corresponding qubits has a
clear physical meaning of the initially provided quantum information.
Discussion of the radiation channel is also interesting, because the
transformation of the initial qubit into the photon field enables a wide choice
of subsequent transformations. A particular question that can be raised here is
how rapidly could the information be recycled after a single use of a
qubit--photon field channel?

\begin{figure}[ht]
\begin{center}
\epsfysize=0.3\textwidth\epsfclipon\leavevmode\epsffile{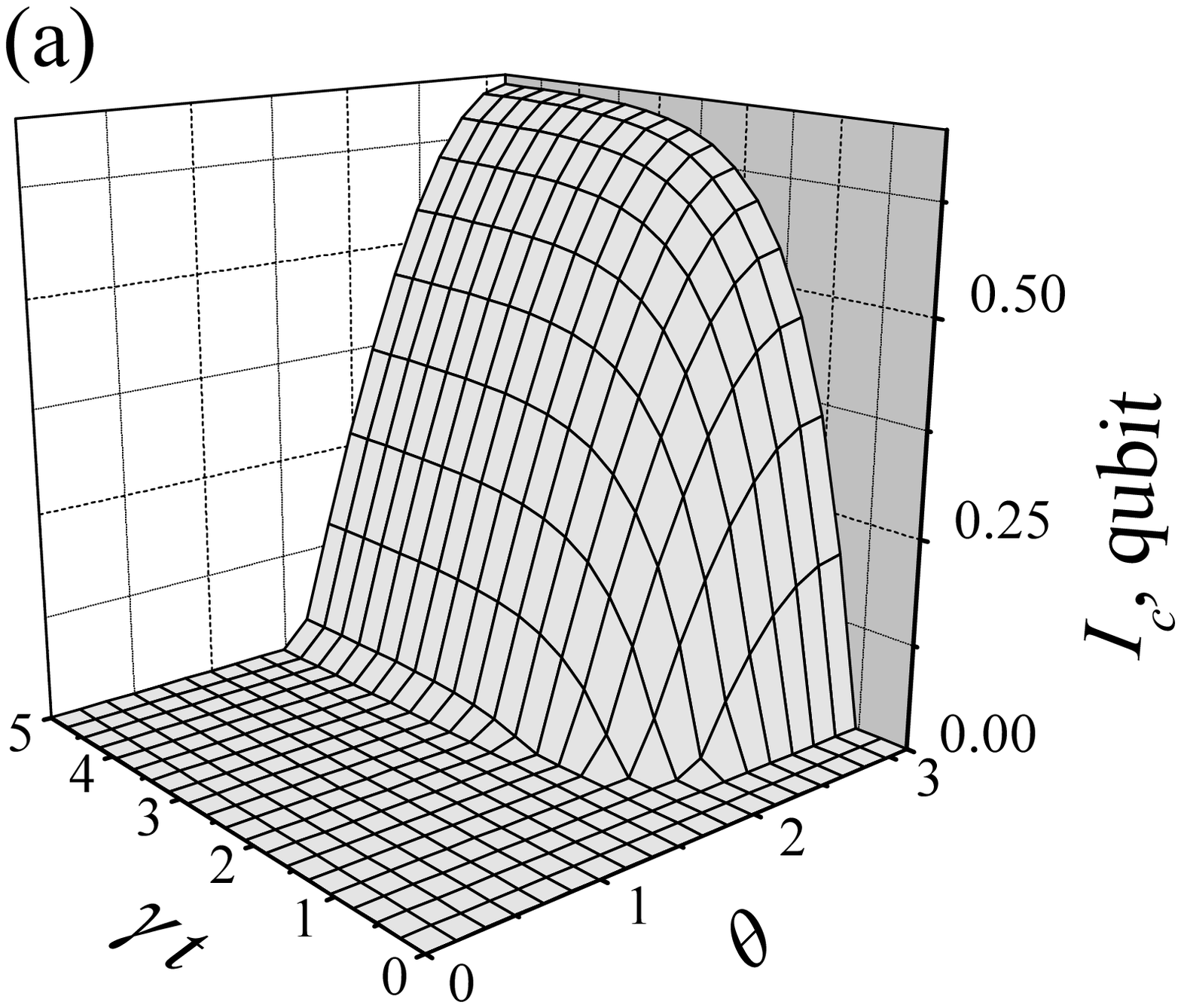}
\hspace{1.5cm}
\epsfysize=0.3\textwidth\epsfclipon\leavevmode\epsffile{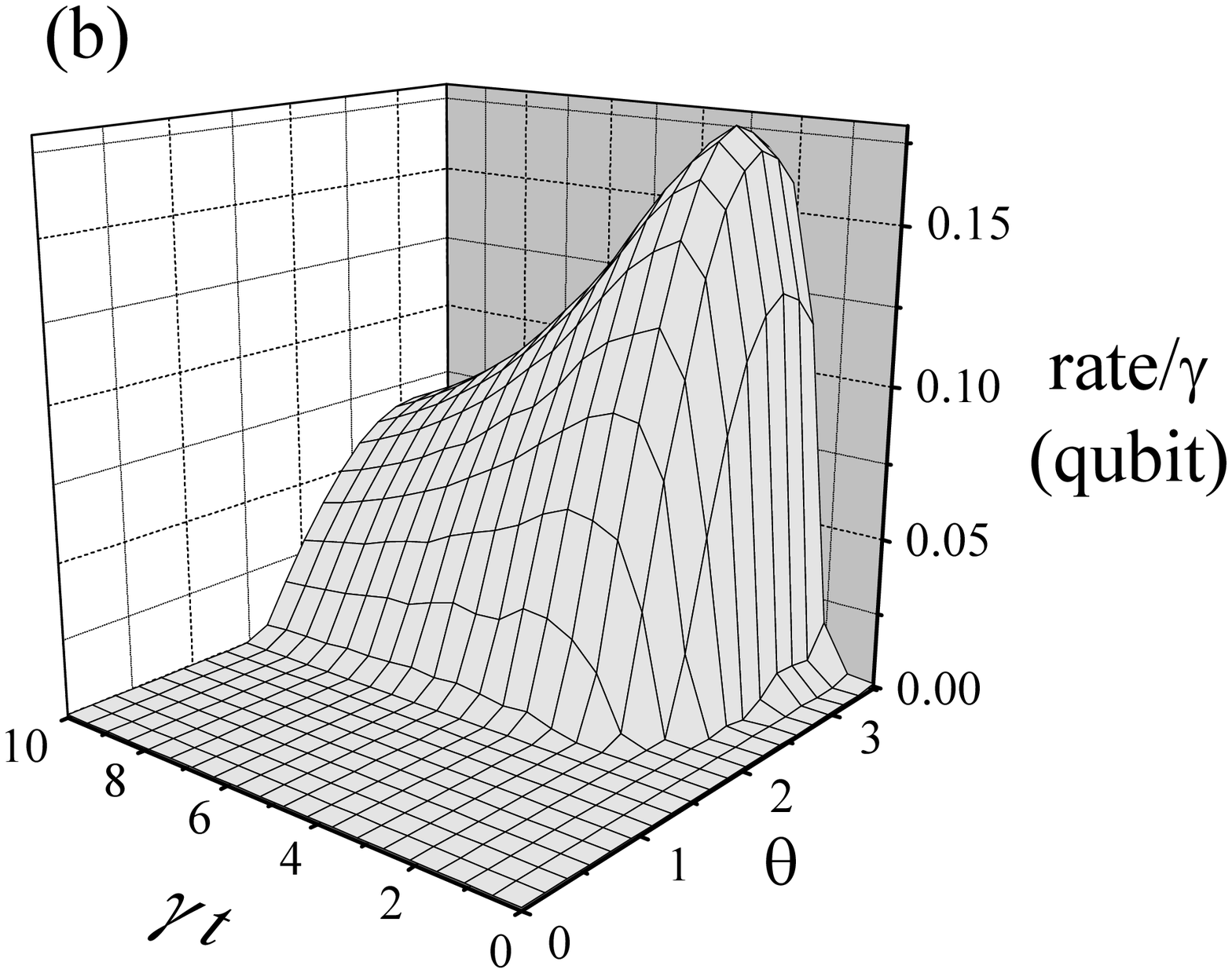}
\end{center}
\caption{a) The coherent information in a symmetric $\Lambda$-system as
a function of dimensionless time $\gamma t$ and action angle
$\theta=\Omega\tau_p$ for the maximum entropy input state; $\gamma$ is
the decay rate, $\Omega$ is the effective Rabi frequency and $\tau_p$
is the exciting pulse duration. b) Dependence of the information rate
on the cycle duration $t$ and action angle $\theta=\Omega\tau_p$.}
\label{fig4}
\end{figure}

Details of the calculations of the coherent information exchange for this
channel are given in Ref.\ \onlinecite{lasphys00}. The dependence of the coherent
information on time and laser field action angle for a symmetric
$\Lambda$-system is shown in Fig.\ \ref{fig4}a for a maximum entropy qubit state
$\hat\rho_A=\hat I/2$, when information does not depend on the individual field
intensities of the two applied laser fields.

It can be easily seen from Fig.\ \ref{fig4}a that there is an optimum value for
the information rate $R=I_c/t$, $t=\tau_c$, if we introduce a periodic use of
the information channel with a cycle duration $\tau_c$, so that after each
cycle the initial state is instantaneously renewed. The calculation results for
the rate $R$ for a symmetric $\Lambda$-system with the partial decay rates
$\gamma_1{=}\gamma_2{=}\gamma$ are shown in Fig.\ \ref{fig4}b \cite{bokarev}.
The total optimum rate is $R_0=0.178\gamma$. Thus, the process of atom--photon
field information exchange sets the corresponding rate limit on using the
coherent information stored in the $\Lambda$-systems. The order of its
magnitude is given by the decay rate of the excited state, while an exact value
depends on the partial decay rates $\gamma_{1,2}$ of the $\Lambda$-system
transitions. At the limit of a two-level radiative system, $\gamma_1=0$ or
$\gamma_2=0$, the optimum rate is equal to $0.316\gamma$.

\section{Compatible information}
\label{section:compatible}

For one-time average values, one can restrict representation of quantum
internal incompatibility in an equivalent form of classical probability
distribution on the quantum states of interest. Then, for the probability
measure
\begin{equation}\label{Pa}
P(d\alpha)=\bra{\alpha}\hat\rho_A\ket{\alpha}dV_\alpha
\end{equation}

\noindent on the space of all quantum states the average value of an operator
$\hat A=\sum\lambda_n\ket{n}\bra{n}$ can be written as $\avr{\hat
A}=\sum\lambda_ndP/dV_{\alpha}(\alpha_n)$, where $\ket{\alpha_n}= \ket{n}$.
Here $dV_\alpha$ is the volume differential in the space of physically
different states of the $D$-dimensional Hilbert space $H_A$ ($\int
dV_\alpha=D$), which, for example, for a qubit system with $D=2$ is the Bloch
sphere (see Sec.~\ref{section:meaning}). Eq.~(\ref{Pa}) is an average of the
projective measure
\begin{equation}\label{Ea}
\hat E(d\alpha)=\ket{\alpha}\bra{\alpha}dV_\alpha,
\end{equation}

\noindent which is a specific case of non-orthogonal decomposition of unit
\cite{gTK73}, or positive operator-valued measure (POVM)\cite{preskill}. POVMs
represent some physical measurement procedures made in a compound space
$H_A{\otimes} H_a$ with an appropriate additional space $H_a$ and joint density
matrix $\hat\rho_A{\otimes}\hat\rho_a$, which gives no additional information
about $A$ beyond the information given by $\hat\rho_A$.

Let us assume that two Hilbert spaces, $H_A$ and $H_B$, of the corresponding
quantum systems $A$ and $B$ and the joint density matrix $\hat\rho_{AB}$ in
$H_A{\otimes}H_B$ are given. Specifically, they can correspond to the
subsystems of a compound system $A{+}B$, given at the same time instant $t$,
or be defined as input and output of an abstract quantum channel of a real
physical system. Described above subsystems $A$ and $B$ are compatible.
Therefore, a joint measurement represented with the two POVMs as $\hat
E_A\otimes\hat E_B$ gives no extra correlations between output and input
measurements and the respective joint input--output probability distribution
takes the form:
\begin{equation}\label{Pinout}
P(d\alpha,d\beta)={\rm Tr}\,\bigl[\hat E_A(d\alpha)\otimes\hat E_B(d\beta)
\bigr]\hat\rho_{AB}.
\end{equation}

\noindent The corresponding Shannon information $I=S[P(d\alpha)]+ S[P(d\beta)]-
S[P(d\alpha,d\beta)]$ defines then the \emph{compatible} information measure
\cite{LP2001}.

The physical meaning of the compatible information depends on the specific
choice of the measurement and represents the quantum information on input
obtainable from the output via the POVMs, which select the information of
interest in the classical form of the corresponding $\alpha$ and $\beta$
variables, the information carriers.

Let us consider the case when $\alpha$ and $\beta$ enumerate all the quantum
states of $H_A$ and $H_B$, in accordance with Eq.~(\ref{Ea}). In this case,
compatible information is distributed over all quantum states and associated
with the internal quantum uncertainty, which is  taken into account in the
distribution (\ref{Pa}). Specifically, quantum correlations due to the
possible entanglement between $A$ and $B$ are taken into account in the
joint probability (\ref{Pinout}). Moreover, the compatible information in
this case yields the \emph{operational invariance} property \cite{zeil},
which is when all the non-commuting physical variables are taken into
account equivalently. Such classical representation of the quantum
information can be associated with the representations of quantum mechanics
in terms of classical variables\cite{Glauber}.

\section{Amount of Information Attainable by an Experimental Setup}
\label{section:setup}

\begin{figure}[ht]
\begin{center}
\epsfysize=0.3\textwidth\epsfclipon\leavevmode\epsffile{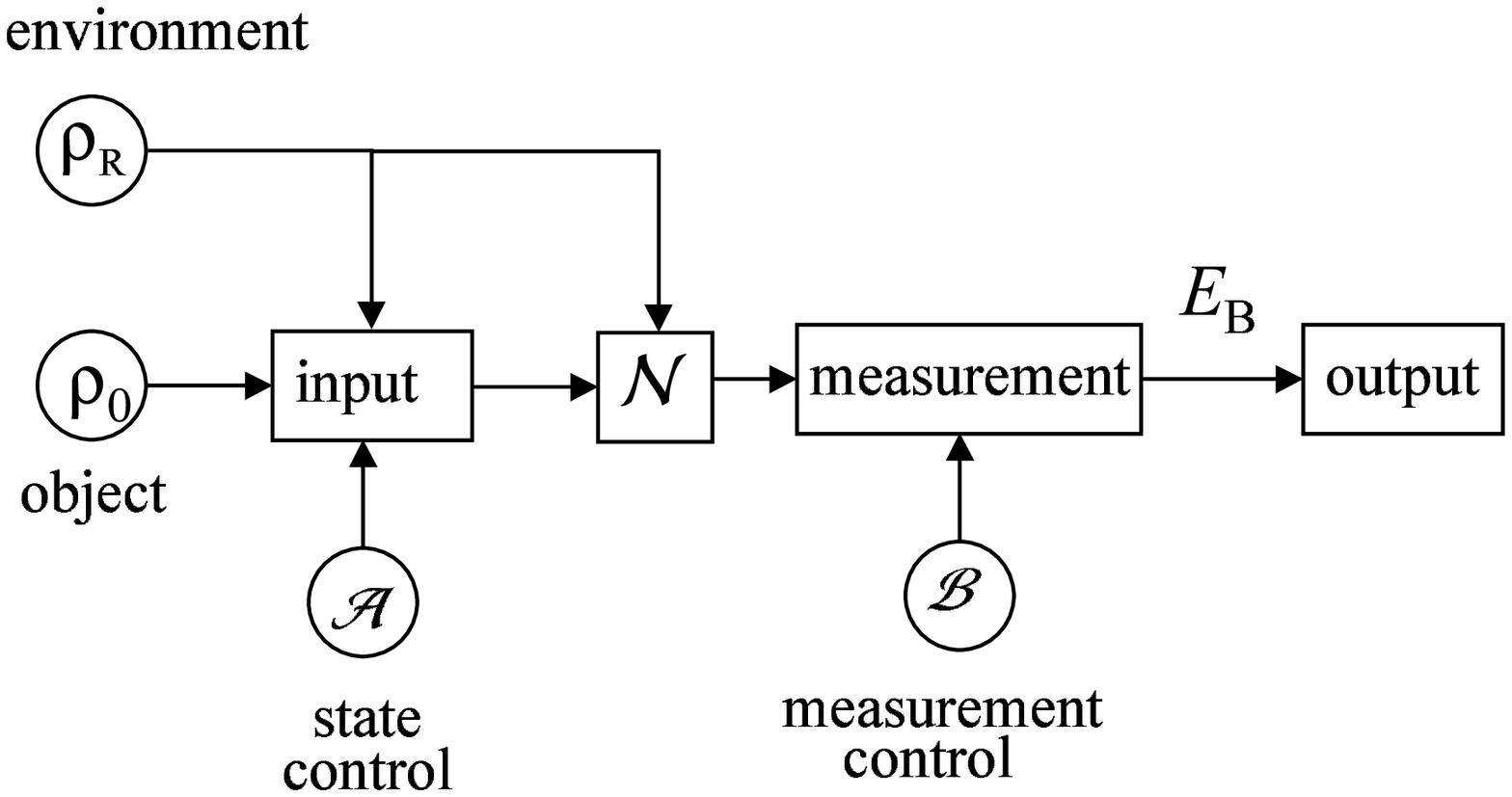}
\end{center}
\caption{Information structure of a quantum experimental setup. An
object accompanied with the noise environment undergoes the state
control interactions, produces the input information ensemble,
depending on either the object dynamical parameters or quantum states
of interest. Then, after the channel superoperator transformation
${\cal N}$ the output information is measured. ${\cal A}$ and ${\cal
B}$ denote transformations provided with the controlling interactions,
$E_B$ stays for the measurement procedure in the form of the
corresponding POVM.} \label{fig5}
\end{figure}

Our previous discussion of the generalized measurements encourages us to
introduce in this section a likelihood mathematical concept of information
attainable by an experimental setup, which certainly is one of the key goals of
the Quantum Information Theory. It is difficult to define the information model
corresponding to the experimental setup under consideration in general form.
Therefore, one has first to specify the input and output information of
interest (which is actually the most difficult point here). We propose here a
solution illustrated by the block scheme shown in Fig.\ \ref{fig5}.

This block scheme corresponds to a typical mathematical structure of a density
matrix of a complex system including two transformations, ${\cal A}$ and ${\cal
B}$, representing control and measurement interactions, correspondingly:
\begin{equation}\label{BNAr}
\hat\rho_{\rm out}={\cal B}{\cal N}{\cal A}\,\hat\rho_{\rm in}.
\end{equation}

\noindent Here $\hat\rho_{\rm in}$ and $\hat\rho_{\rm out}$ are the initial and
final density matrices for the degrees of freedom, chosen in a mathematical
model of the experimental setup. Superoperators ${\cal A}$, $\cal B$, and $\cal
N$ are associated with the preparation of the information, the measurement, and
the transmission of the information to the output, correspondingly. This
markovian-type structure is not the most general one---for simplicity we assume
that the reservoirs corresponding to each transformation are independent and
their density matrices can be separated from $\hat\rho_{\rm in}$. Only under
this simplification we can get a separated combination of the three
superoperators and the input density matrix and, as a result, get a relatively
simple mathematical representation of the information structure in terms of the
corresponding decompositions of $\cal A$ and $\cal B$. Still, we have to keep
in mind that a proper generalization of Eq.~(\ref{BNAr}) may be necessary in a
general case.

Preparation of the information always involves some interactions, resulting in
the corresponding transformations, which are unitary only if all the involved
degrees of freedom are taken into account. We have to include also interaction
with the reservoir represented with a non-unitary superoperator. We will
discuss here the recepies for two possible choices of a physical information of
interest:
\begin{description}
\item[(i)] the system dynamic parameters $a$,
\item[(ii)] the system dynamic states $\ket a$.
\end{description}

For the choice (i), the required information goal can be achieved with the use
of the dynamical evolution operators $U_A(a)$, which in its turn may depend on
the controlling parameters $c$. \emph{A priory} information on $a$ is included
in a proper chosen probability measure $\mu(da)$. Corresponding superoperator
${\cal A}$ is then can be written as ${\cal A}=\int{\cal A}_a\mu(da)$ with
\begin{equation}\label{Aa1}
{\cal A}_a=\avr{U_A^{}(a)\odot U_A^{-1}(a)}_E,
\end{equation}

\noindent where symbol $\odot$ denotes the place to substitute with the
transformed density matrix and brackets denote averaging over the noise
environment.

For the choice (ii), the required information goal can be achieved with the use
of the measurement superoperator transformation composed of superoperators
\begin{equation}\label{Aa}
{\cal A}_a=\avr{\ket{a}\bra{a}\odot\ket{a}\bra{a}}_E.
\end{equation}

\noindent The corresponding sum ${\cal A}=\sum{\cal A}_a$ is the measurement
superoperator represented with an averaged standard decomposition $\sum_i\hat
A_i\odot\hat A_i^+$ of the completely positive trace-preserving superoperator
\cite{kraus83} with a properly specified operators $\hat A_i=\hat
A_i^+\to\ket{a}\bra{a}$. Keeping in mind that $a$ can represent a continuous
variable, we have to use a generalized representation ${\cal A}= \int{\cal
A}_a\mu(da)$ in the integration form with a proper measure $\mu(da)$, providing
a corresponding decomposition of unit (POVM) $\int \ket{a}\bra{a}\mu(da)=\hat
I$.

In most general form, the superoperator sets (\ref{Aa1}), (\ref{Aa}) are
represented with an arbitrary {\em positive superoperator measure} (PSM) ${\cal
A}(da)= {\cal A}_a\mu(da)$, which is a decomposition of a completely positive
trace-preserving superoperator. PSM satisfies the conditions of complete
positivity, ${\cal A}(da)\hat\rho \geqq0$, and normalization, ${\rm Tr}
\int\!{\cal A}(da)\hat\rho=1$. The latter can be expressed in an equivalent
form of preservation of the unit operator $\int{\cal A}^*(da)\hat I=\hat I$ by
the conjugate PSM ${\cal A}^*$.

It is worth to discuss here a special case when the POVMs are represented by
Eq.~(\ref{Ea}) with all the states of the Hilbert spaces $H_A$ and $H_B$
corresponding to ${\cal A}$ and ${\cal B}$, again. This definition of the POVMs
restricts the information attainable by an experimental setup due to the basic
physical limitations underlying the chosen mechanism of obtaining quantum
information. The latter is represented here in a ``solid" classical form
enabling its copying and free use. This property may as well be assigned by
default to the meaning of the term ``information", by contrast to the opposing
meaning of the coherent information discussed in
Secs.~\ref{section:meaning}--\ref{section:rate}.

Repeating the above argumentation for the measurement superoperator ${\cal B}
=\int{\cal B}(db)=\int{\cal B}_b\nu(db)$ with ${\cal B}_b$ in the form of
Eq.~(\ref{Aa}), we can implement the input and output information in the form
of classical variables $a$ and $b$ for both choices, (i) and (ii), of the
information of interest. The corresponding joint probability distribution is
then given by
\begin{equation}\label{Pab}
P(da,db)={\rm Tr}\,{\cal B}(db){\cal N}{\cal A}(da)\,\hat\rho_{\rm in}.
\end{equation}

\noindent This distribution is always positive and normalized to 1. It gives an
experimenter the statistical correspondence between the states of interest and
output information attainable by the experimental setup. The corresponding
\emph{information capacity} of the setup can be expressed in the quantitative
form as the responding Shannon information, which then can be used for
optimization of the setup parameters.

It is important to note that mutual compatibility of the $\ket a$ and $\ket b$
states for (i) choice is not declared here and, in general case, the states can
correspond to the non-commuting projectors. In a trivial extreme, they could be
the same states and all the information is sent with zero error probability. If
the states belong to the different physical subsystems, they may carry on
quantum correlations due to the corresponding structure of the channel
superoperator ${\cal N}$. A simplest example could be given by ${\cal
N}=U_{AB}\odot U_{AB}^{-1}$ with $U_{AB}$ being the entangling unitary
transformation.

The control parameters $c$ may be either fixed or be set of used values
$c\in\mathbb{C}$. For their optimization one can use the Shannon information
measure. The unknown probability distribution $\mu(da)$ of the dynamical
parameters $a$ for the case (i) can be calculated in terms of the classical
decision theory\cite{Wald} and no quantum mechanics is necessary. As for the
specification of the action ${\cal B}_b$ of the measurement system in the form
(\ref{Aa1}), it may be generalized in the form of a general type PSM. Two PSMs
${\cal A}(da)$ and ${\cal B}(db)$ cover a wide range of state control and
measurement systems implemented into the model of the experimental setup.

\section{Conclusions}
\label{section:conclusions}

In the paper we classified the quantum information into the \emph{classical,
semiclassical, coherent,} and \emph{compatible} information based on the
compatibility property. This list exhausts all basically different types of
quantum information.

Physical meaning of the coherent information is an amount of the internal
incompatibility exchanged between two systems and measured as an entanglement
\emph{preserved} between the output and the reference system. Introduced here
\emph{one-time coherent information} sets a correct correspondence between the
Schumacher's and modified Stratonovich's approaches. We calculated the coherent
information exchange {\em rate} of a $\Lambda$-system via photon field that
does not exceed $0.178\gamma$ for a symmetric $\Lambda$-system and
$0.316\gamma$, otherwise.

We introduce here for the first time the \emph{compatible information}, which
is an adequate characteristic of the quantum information exchange between
compatible systems. The compatible information can be expressed in terms of
classical information despite internal incompatibility, by contrast with the
coherent information, which is basically irreducible to the classical terms.

It is shown that internal compatibility of the input and output quantum
information seems an adequate restriction for a physical information in an
experimental setup. It makes possible quantitative characterization of the
available \emph{information capacity} of the experimental setup. Then,
information exchange between the subsystem, preparing information, and the
measuring device is formulated as a probabilistic correspondence between the
classical variables determining the corresponding dynamical evolution and the
measured output values. A general mathematical representation of the
information generation and its readout is presented in the form of two PSMs.
This representation of physical information exchange in an experimental setup
seems to be promising in direct application of Quantum Information Theory to
the demands of experimental physics.

\acknowledgments This work was partially supported by RFBR grant no.
01--02--16311, the State Science-Technical Programs of the Russian Federation
``Fundamental Metrology" and ``Nano-technology", and INTAS grant no. 00--479.

\end{document}